\documentclass[12pt,preprint]{aastex}
\begin{document}

\title{The Effect of Star Formation History on the Inferred Initial
Stellar Mass Function}

\markboth{\sc Elmegreen \& Scalo}{\sc Star Formation History and the IMF}

\author{Bruce G. Elmegreen \affil{IBM Research Division, Yorktown Heights,
New York 10598; email: bge@watson.ibm.com}}

\author{John Scalo \affil{Department of Astronomy, University of Texas,
Austin, Texas 78712; e-mail:parrot@astro.as.utexas.edu}}


\begin{abstract}
Peaks and lulls in the star formation rate (SFR) over the history
of the Galaxy produce plateaux and declines in the present day
mass function (PDMF) where the main-sequence lifetime overlaps the
age and duration of the SFR variation. These PDMF features can be
misinterpreted as the form of the intrinsic stellar initial mass
function (IMF) if the star formation rate is assumed to be
constant or slowly varying with time. This effect applies to all
regions that have formed stars for longer than the age of the most
massive stars, including OB associations, star complexes, and
especially galactic field stars. Related problems may apply to
embedded clusters. Evidence is summarized for temporal SFR
variations from parsec scales to entire galaxies, all of which
should contribute to inferred IMF distortions. We give examples of
various star formation histories to demonstrate the types of false
IMF structures that might be seen. These include short-duration
bursts, stochastic histories with log-normal amplitude
distributions, and oscillating histories with various periods and
phases. The inferred IMF should appear steeper than the intrinsic
IMF over mass ranges where the stellar lifetimes correspond to
times of decreasing SFRs; shallow portions of the inferred IMF
correspond to times of increasing SFRs.  If field regions are
populated by dispersed clusters and defined by their low current
SFRs, then they should have steeper inferred IMFs than the
clusters. The SFRs required to give the steep field IMFs in the
LMC and SMC are determined. Structure observed in several
determinations of the Milky Way field star IMF can be accounted
for by a stochastic and bursty star formation history.
\end{abstract}

\section{Introduction}

The star formation rate (SFR) in galaxies varies over a wide range
of timescales. The ratio of recent to past-average SFR (denoted
$b$; see Scalo 1986) highlights the most recent changes. This
ratio has been estimated using the equivalent width of H$\alpha$
(Kennicutt, Tamblyn \& Congdon 1994; Kennicutt 1998) and the ratio
of far-infrared to blue luminosity (Tomita, Tomita, \& Saito
1996), both of which suggest variations of up to an order of
magnitude in galaxies of the same Hubble type. Related evidence
for repetitive near-global bursts can be inferred from lopsided
galaxies (Rudnick, Rix, \& Kennicutt 2000), discrepancies between
different SFR diagnostics (Sullivan et al. 2001), pixel-by-pixel
models of disk galaxies at moderate redshifts (Glazebrook et al.
1999), and systematics of Tully-Fisher residuals for local
galaxies (Kannappan, Fabricant, \& Franx 2002). Recent work using
much larger samples supports the idea that global variations are
common in most galaxies: Salim et al. (2005) found SFR variations
using GALEX UV and SDSS optical properties of $6500$ galaxies, and
Brinchmann et al. (2004) found them using optical emission lines
in $10^5$ SDSS galaxies. Related variations have been found in the
H$_2$-to-HI ratio (Keres, Yun \& Young 2003) and mass-to-light
ratio (Roberts \& Haynes 1994) for each Hubble type, and in the
SFR for a given gas column density (Kennicutt 1998).

Population synthesis studies of nearby dwarf Irregulars also show
a history of intense variations (Hunter 1997; Tolstoy et al. 1998;
Hodge 1989; Carraro 2002; Grebel 2001; Van Zee 2001; Grebel \&
Gallagher 2004), while BCD galaxies are bursting today (Searle \&
Sargent 1972; Kunth \& Ostlin 2000; Hopkins, Schulte-Ladbeck, \&
Drozdovsky 2002). Bursty SFR behavior in low-mass galaxies was
found in the FORS Deep Field study of galaxies at redshift 1.5
(Bauer et al. 2005). The cause of most of these SFR variations is
unknown, although some can be ascribed to tidal interactions
between galaxies.

There is also evidence for SFR variations in the Milky Way. Stars
in the solar neighborhood with ages up to several Gyr show SFR
variations that have been interpreted as periodic or irregular
(Barry 1988; Majewski 1993; Rocha-Pinto et al. 2000a,b; Hernandez,
Valls-Gabaud, \& Gilmore 2000; de la Fuente Marcos \& de la Fuente
Marcos 2004). Azimuthal orbit diffusion for ages more than a few
$10^8$ yr suggests these stars sample an entire annulus of our
Galaxy (see Rocha-Pinto et al. 2000a). Thus the variations may be
global.

Star formation events triggered by spiral density waves, swing
amplified instabilities, and superbubbles have a timescale of
$\sim10^8$ yrs. Gould's Belt is an example of a local burst
(P\"oppel 1997). Another local burst $2-4\times 10^8$ yr ago was
inferred from chromospheric ages (Barry 1988; Rocha-Pinto et al.
2000a,b), a bump in the present-day luminosity function (Scalo
1987), and a feature in the white dwarf luminosity function (Noh
\& Scalo 1990).

Compression and collapse from interstellar turbulence (see review
in MacLow \& Klessen 2004) and compression from the pressures of
massive stars should lead to bursty star formation on a wide
variety of spatial and temporal scales, including the small scales
of cloud cores and clusters. For example, Hillenbrand (1997) and
Palla \& Stahler (2000, 2002) found an acceleration of SFRs in
local star-forming regions over the last $\sim10^7$ yr, while
Selman et al. (1999) found three large bursts in 30 Dor. A
deceleration should occur when an active cloud core gets
dispersed.

The purpose of this paper is to show the effect of these temporal
variations on the present day mass function of stars, and to
illustrate how the stellar initial mass function (IMF) can be
erroneously derived if the detailed star formation rate history
(SFH) is not properly included.  The argument applies to all
regions that have formed stars longer than the age of the most
massive stars. Thus it applies to the inferred IMFs of OB
associations, star complexes, very young clusters, and galactic
field stars. It should also apply to the IMFs derived from
luminosity functions inside young embedded clusters if the SFR
varies during cluster formation. In these cases the SFH and IMF
should be determined simultaneously, which requires ages and
masses for stars from comparisons with isochrones (e.g.
Hillenbrand 1997; Selman et al. 1999; Hillenbrand \& Carpenter
2000; Muench et al. 2003). An additional complication occurs if
the SFH depends on mass, i.e. if the massive stars form in
different cloud cores than the low mass stars (e.g.,
DeGioia-Eastwood et al. 2001; see Elmegreen 2004).

The inferred IMF of an old cluster should not depend on the
history of its formation if the most massive stars remaining in
the cluster have main sequence lifetimes that are older than the
duration of star formation in the cluster.  Aside from evaporation
and ejection, old clusters should show their correct intrinsic
IMFs up to the most massive star that is still on the main
sequence.  The highest-mass part of the cluster IMF, which is now
lost to stellar evolution, could have shown SFH-related deviations
at an earlier stage if the cluster were observed then.

The global properties of galaxies such as integrated light and
chemical abundances are also affected by the SFH. If large-scale
bursts are as common as indicated, then models ignoring these
bursts will derive the wrong PDMF even if the correct IMF is used.
On the other hand, if there exists an intrinsic universal IMF
whose form can be derived from open cluster studies (see Scalo
1998, 2005; Kroupa 2002), then our results imply that the SFH can
be inferred from the PDMF. This technique was used by Scalo (1987)
for the Milky Way, and Mighell \& Butcher (1992) for the Carina
dwarf spheroidal galaxy.

\section{Inferred IMFs}

The SFH,
$S(t)dt$, combined with the intrinsic IMF, $f(M)d\log M$, gives the
PDMF $\phi(M)d\log M$,
\begin{equation}\phi(M)=f(M)\int_0^{t(M)} S(t)
dt\label{eq:pdmf}\end{equation} where $t(M)$ is the main-sequence
lifetime of a star of mass $M$. In this equation, the age $t$
increases into the past from the present time at $t=0$. We assume
$f(M)$ is independent of time and measured in log-intervals of
stellar mass.  Only the relative variations in the star formation
rate, $S(t),$ have an effect on the IMF; the absolute SFR does not
matter. Thus the IMF modulations we discuss here apply to all
regions where stars have formed with a relative rate of $S(t)$ and
a time-invariant intrinsic IMF $f(M)$ of any type. We do not
consider time-variable intrinsic IMFs in our examples, but such
variations could be present in galaxies as well.

The PDMF is not observed directly but inferred from the luminosity
function using a mass-luminosity relation. This procedure can be
used only for groups of stars in the main sequence phase of
evolution: the mass-luminosity relation is not one-to-one for
pre-main sequence stars.  Usually the field or galactic average
IMF is derived from the PDMF by assuming a SFH that varies
smoothly over large timescales. For our example, a constant SFH,
$S_0$, would give an inferred IMF
$f_{infer}(M)=\phi(M)/\left[S_0t\left(M\right)\right]$. For the
approximation $t(M)\sim M^{-2.2}$ (Mihalas \& Binney 1981), this
gives $f_{infer}(M)\propto\phi(M)M^{2.2}$.

Structure in the inferred IMF comes from unknown variations in
$S(t)$. This structure may be derived analytically for simple
burst cases of $S(t)$, which include a sudden increase followed by
an exponential or power-law decline, or a sinusoidal variation.
These analytically solvable cases can be used to show that the
amplitude and mass range of the resulting features in $f_{infer}$
are determined primarily by the age, duration, and amplitude of
the burst in $S(t)$. The shapes of the features are relatively
insensitive to the precise form of $S(t)$ for these burst cases.

In the models below, the intrinsic IMF is assumed for illustrative
purposes to be $f(M)\propto M^{\Gamma}$ for $M>1$ M$_\odot$ with
$\Gamma=-1.5$, a value that brackets empirical estimates
($\Gamma=-1.35$ is the Salpeter IMF slope; see review in Chabrier
2003). Different $\Gamma$ introduce different tilts in the
inferred IMF without changing the form of the IMF distortions. The
IMF below 1 M$_\odot$ is not included because such stars live
longer than $10^{10}$ yr, the longest age considered here. The
conversion between mass and main sequence lifetime is assumed to
be given by the fit to the models of Schaller et al. (1992) in
Reid, Gizis \& Hawley (2002):
\begin{equation}\log_{10}\left(t\right)=
10.015-3.461\log_{10}\left(M\right)+0.8157\log_{10}\left(M\right)^2
.\end{equation}.

Figure 1 shows the effect of a delta-function burst of star
formation on the inferred IMF if $S(t)$ is erroneously assumed to
be constant. On the left is the dimensionless SFH (with each case
displaced upward for clarity), and on the right is the inferred
IMF made under this erroneous assumption.  The actual SFH is taken
to be a constant value of 1 except in bursts where it increases by
a factor of 10. Features in the inferred IMFs occur at each mass
where the main-sequence time equals the burst time. If the IMF is
divided by the intrinsic IMF, $M^{\Gamma}$, and the derivative
taken, then the SFH will be recovered (see Eq. 5 below). Note that
the inferred IMF is exactly the intrinsic IMF at masses above the
main-sequence turn off mass for the burst age because these
massive stars were not around at the time of the burst. The IMF
flattens just below the turnoff mass because some of these stars
formed in the burst. For bursts with a finite duration, the sharp
drop in the inferred IMF above the turnoff mass would be
broadened.

The inferred IMF is plotted up to 100 M$_\odot$ in all of the
figures even though the galactic IMF is not well known above
$\sim50$ M$_\odot$ and may not be a continuation of the assumed
power law from lower masses.  In addition, the highest-mass
stellar lifetimes are comparable to the shortest timescales for
variations considered here, so there are no features in any of our
inferred IMFs above $\sim50$ M$_\odot$.  Nevertheless, the
plotting is done in this way to give a high-mass baseline for
viewing the IMF variations at lower mass.

Figures 2-5 show other inferred IMFs for various star formation
histories. Figure 2 has dimensionless SFHs that are made from the
exponentiation of the Fourier transform of noise that was
multiplied by $k^{-\beta}$ for wavenumber $k$. We make these
stochastic SFHs in the following way (see also Weinberg \& Cole
1992 for cosmological applications and Elmegreen 2002 for
applications to cloud mass functions). We first generate 5000
random numbers $R_1$ and another 5000 $R_2$ with values uniformly
sampled between $-0.5$ and $0.5$. Then we take the Fourier
transform of $k^{-\beta}$ times these random numbers, using
integers for $k$ between 1 and 5000 and integers for time $t$
between 1 and $10^4$. The units of $t$ are My. This gives
\begin{equation}
F(t)=\sum_{k=1}^{5000} k^{-\beta}\left(R_1(k)\sin\left[\omega
kt\right]+ R_2(k)\cos\left[\omega kt\right]\right).
\end{equation}
Finally, we take for the SFH the exponentiation of this result:
\begin{equation}
S(t)=\exp\left(F\left[t\right]\right)
\end{equation}
The frequency is $\omega=2\pi/\left(10^4 My\right)$. The
multiplication of the noise by $k^{-\beta}$ for $\beta=0.25$, 0.5,
and 1 makes the power spectrum of $F(t)$ a power-law with a slope
of $-0.5$, $-1$, and $-2$, respectively. The amplitudes of $F(t)$
have a Gaussian pdf, so the exponentiation of $F(t)$ gives $S(t)$
a log-normal pdf. Because transformation of the one-point pdf can
distort the power spectrum, we checked the power spectrum before
and after exponentiation and found little difference aside from an
amplitude shift, and therefore did not apply the correction
described by Weinberg \& Cole (2001). The corresponding SFHs are
plotted in the bottom, middle, and top panels, respectively. The
rms dispersion of $F(t)$ is normalized to 1 in each case (not
shown in the equations above), and the average value of $F(t)$ is
normalized to 0.1 for the lower SFH in each panel, 1 for the
middle, and 10 for the upper. Note that the vertical scales differ
for each curve in a panel; the normalization of the average $F(t)$
does not affect the inferred IMF. As a result of these steps,
$S(t)$ has a log-normal pdf and an approximately power-law power
spectrum. The fat tail of the log-normal pdf provides the
intermittency that is meant to mimic the short bursts and longer
lulls expected from turbulent density variations in the ISM.

Figure 3 shows 11 different random-number renditions of the same
stochastic history model used for the middle curve in the middle
panel of Figure 2. That is, the power-law slope in Figure 3 is
$-1$ and the normalized average of $F(t)$ is 1. The inferred IMFs
vary a lot from curve to curve, reflecting the different details
in the SFHs plotted on the left.  These variations would be
greater or smaller for SFHs with power spectra of -2 or -0.5,
respectively, as can be inferred from Figure 2. Stochastic effects
like this could be responsible for some of the structure seen in
the field star initial mass function when the range of masses is
large and the IMF features are not smoothed by large mass bins.
Examples are the ledges and dips in the IMF by Scalo (1986, Fig.
16), Rana (1991, Fig. 3), and Basu \& Rana (1992, Fig. 6).
Structure can even be seen in Salpeter's (1955) IMF.   Of course
these features could also be due to structure in the functions
needed to transform from the PDMF to the IMF (see Scalo 1986), or
they could be intrinsic to the IMF. Unfortunately, the most recent
determinations of the field star IMF (Reid, Gizis, \& Hawley 2002;
Schroder \& Pagel 2003) do not cover a large enough range in mass
above 1 M$_\odot$ for such structure to be noticed. Future studies
of the intermediate-mass IMF that include a larger mass range
should find the predicted signatures of SFH variations.

Figure 4 has sinusoidal SFHs with various phases shown on the
left. Each sine function has a constant added so that zero is the
minimum value of the star formation rate in the bottom two left
panels, with a full range from 0 to 2, and 0.5 is the minimum
value in the top left panel, with a range from 0.5 to 1.5. This
top panel in Figure 4 shows the effect of diluting the sinusoidal
variation with a constant minimum SFR. The bumps in the IMF on the
right are correspondingly weaker for this diluted case than they
are in the bottom two panels. This difference illustrates how the
relative amplitude of the SFH variations control the magnitude of
the IMF effect.

Figure 4 indicates that the intermediate-to-high mass part of the
IMF can become very steep if the SFH declines continuously to the
present day.  This is the case in the middle panel for the sine
function having a value of 0 at $t=0$ (the blue curve).  The
figure also shows how the IMF is hardly affected below the mass
whose main sequence lifetime equals the period of the sinusoidal
variation of the SFH. This is because lower mass stars form during
many periods of the variation so the ups and downs of the SFH
average out to a nearly steady star formation rate.

In Figure 5, the inferred IMFs (right hand side) and IMF slopes at
5, 10, and 20 M$_\odot$ (top left) are plotted versus the phase
from the sinusoidal model in the lower left. The phase is measured
relative to $2\pi$, which is a full cycle of the sinusoid. The
inferred IMF is distorted in a way that depends on the current
phase in the star formation variation. An example was previously
noted for Figure 4: if the most recent SFR has declined sharply to
a value of zero today (as for the green curve in the lower left
panel of Fig. 5), then the inferred IMF would have a relatively
steep slope from intermediate to high mass (corresponding to the
green and lower curve in the right-hand panel). Or, if the SFR had
its most recent minimum 25 My ago and then started to rise again
(red curve on the lower left), then the inferred IMF would have a
dip at the mass whose main-sequence lifetime is 25 My (red line on
the right). The phase of the sinusoid in the first example is 0,
and the IMF slopes at 5, 10, and 20 M$_\odot$ are $-3.8$, $-5.0$
and $-3.9$, as shown in the top left panel for phase$=0$. The
phase in the second example is $1/8$ of $2\pi$, and the IMF slopes
at these three masses are $-6.5$, 0.3, and -1.09, respectively, as
shown in the top left for phase$=0.125.$  In general, the time of
the most recent significant dip in the SFR should correspond to
the stellar main sequence life time at the mass where there is a
corresponding dip in the IMF.

Star formation histories like these are not unreasonable. In a
galaxy with irregular spiral structure and star formation
triggered stochastically by turbulent compression, one might
expect the star formation rate to vary with a power spectrum that
is related to the power spectrum of the local dynamical rate. If
the local dynamical rate is proportional to the square root of
density, and the density has a log-normal pdf as in isothermal
turbulence simulations (e.g., Li, Klessen \& Mac Low 2003), then
our SFH in Figure 2 follows.  Lower $\beta$ corresponds to
relatively more short-period SFR variations, which shifts the
irregularities in the inferred IMF toward higher mass while
averaging over and reducing the IMF variations at lower mass.

In a galaxy with global periodic triggering or organization of
star formation by a spiral density wave, the SFH can have periodic
variations. The periods used in Figure 4 are 20 My for the bottom
panel and 200 My for the middle and top panel. The short period
hardly shows up in the inferred IMF because most stars live longer
than that. The long period, which is comparable to the time
between spiral arm passages at typical mid-disk positions, gives a
clear signature in the inferred IMF. This signature can make the
intermediate to high mass slope steeper than the intrinsic slope
if the surveyed stars are down stream from the arm and the SFH has
recently declined (Fig. 5). There is no symmetry for stars
upstream. Far upstream the field stars reflect the previous arm
passage. At the point where the SFR strongly increases, most star
forming regions will still be young and in clusters.

\section{Steep IMFs Inferred for the Field Regions of the LMC
and SMC}

The figures illustrate what happens in regions of galaxies where
the SFR has varied. The observed population is always a PDMF, so
equation 1 has to be applied to obtain the IMF. If, for example, a
decline in the SFR is not included in the derivation of the IMF
from the PDMF, then the inferred IMF will be relatively steep down
to the mass where the turnoff age equals the age at the beginning
of the decline.

The remote field regions of the LMC and SMC may be examples. They
have low SFRs compared to their recent past or else they would be
still be OB associations.  Thus, one might expect the inferred
IMFs to be steeper than the intrinsic IMF if the SFR is
incorrectly assumed to be constant. This could explain some of the
effect found by Massey and collaborators (e.g., Massey et al.
1995; Massey 2002) who assumed a constant SFR for LMC and SMC
fields and derived $\Gamma\sim-4\pm0.5$ in the mass range 25--120
M$_\odot$. A decline in the SFR over the last $\sim5$ My could
explain this $\Gamma$ (using Eq. 2) even if the intrinsic IMF had
the Salpeter slope when the observed stars formed.  The question
is, how large does the decline have to be and is such a decline
reasonable for the whole LMC?

The steep slope found by Massey et al. is steeper than that
obtained for the field regions of the LMC by other groups, and in
the Massey et al. studies it primarily applies to the highest
masses. For example, Massey (2002) divided his star counts into 4
mass bins in Table 12: for the two lowest bins, covering masses
from 25-60 M$_\odot$, the inferred IMF slope is
$\Gamma=\Delta\log\xi/\Delta\log(M)=-2.2$; for the two
intermediate mass bins, covering 40-85 M$_\odot$, $\Gamma=-6.2$,
and for the two highest mass bins, covering 60-120 M$_\odot$,
$\Gamma=-2.6$.  Evidently, the overall $\Gamma\sim-4$ is steep
because of the very steep drop between 40 and 85 M$_\odot$. The
statistical uncertainty in this result is large, however; for the
LMC there are only 16 stars in the two highest mass bins, while
there are 835 stars in the 2 lowest mass bins; for the SMC, which
has the same $\Gamma_{infer}$ overall, there are 5 and 284 stars
in these bins, respectively. Parker et al. (1998) obtained
$\Gamma\sim-1.8\pm0.09$ for the 7-35 M$_\odot$ mass range among
26713 LMC field stars observed with the Ultraviolet Imaging
Telescope, suggesting that steepening at smaller masses is
marginal. The statistical uncertainty in the Parker et al. result
is remarkably low; they assumed continuous star formation in
converting the luminosity function to an IMF.

The PDMF for the LMC field observed by Gouliermis et al. (2005) is
$\propto M^{-6}$ at low mass (0.9-2 M$_\odot$) and $M^{-3.7}$ for
the whole mass range (0.9-6 M$_\odot$).  They comment that the
PDMF for the lowest mass range should equal the IMF because there
should be little evolution at such low masses. In fact the
lifetime of a 2 M$_\odot$ star is only 1 Gy according to equation
2, so there is still time for the SFH to be important if the LMC
disk has an age comparable to that of the Milky Way, $\sim10$ Gy.
Then the PDMF should be divided by $t(M)$ to get the inferred IMF.
For the lower mass range, $t(M)\sim M^{-3.3}$ using equation (2),
and if the SFH is uniform, the resulting IMF is $\propto
M^{-2.7}$. The full mass range should be corrected for $t(M)\sim
M^{-2.9}$ from equation (2), giving an IMF $\propto M^{-0.8}$. The
Gouliermis et al. result is interesting because it suggests the
steep IMF inferred for the field (under the assumption of a
constant SFR) may continue in some regions all the way down to
$\sim1$ M$_\odot$.

These studies allow for the possibility that the LMC and SMC field
IMFs at masses above $\sim40$ M$_\odot$ are steeper than the field
IMFs at intermediate masses, and that both of these are steeper
than what is found in LMC clusters and associations (e.g. Kroupa
2002; Sirianni et al. 2000; Parker et al. 2001; Gouliermis et al.
2005). The implication that dense regions systematically form
proportionally more massive stars than low-density regions was
explored by Elmegreen (2004). However, the results of the present
paper suggest a new explanation for steep field IMFs that is
independent of physical processes during star formation. In this
explanation, the field stars are dispersed clusters or other stars
with an {\it intrinsically} shallow (cluster-like) IMF, but they
occur in places where the SFR has been declining for a time equal
to the lifetime of the lowest mass star on the steep part of the
{\it inferred} IMF.

Figure 6 shows various constraints on the SFH required to produce
a steep inferred IMF from this model alone. It was obtained by
inverting equations (1) and (2). This was done in two steps: first
the PDMF, $\phi(M)$, was recovered from the inferred IMF,
$f_{infer}$, and stellar lifetime function, $t(M)$, using a
constant SFR, $S_0$, as assumed by Massey (2002) and others:
$\phi(M)=f_{infer}(M)S_0t(M)$. Then equation 1 was rewritten as
$\phi(M)/f(M)=\int_0^{t(M)}S(t)dt$ for intrinsic IMF $f$ and
differentiated on both sides with respect to $M$. After
rearrangement, this gives
\begin{equation}
{{S(t)}\over{S_0}}={{d\left(f_{infer}t/f\right)/dM}\over {dt/dM}}
={{f_{infer}}\over{f}}+{{d\left(f_{infer}/f\right)/dM}\over {d\ln
t/dM}}.\label{eq:massey}
\end{equation}
The figure uses an intrinsic IMF with $\Gamma=-1.35$ (the Salpeter
IMF) and three values of an inferred IMF slope
$\Gamma_{infer}=-4$, $-3$, and $-2$. The Salpeter slope was used
for illustration; the results would be similar for other
power-laws. For each of these $\Gamma_{infer}$, two SFHs are
given, one decreasing continuously over the past 10 Gy (dotted
line), which gives $\Gamma=\Gamma_{infer}$ for all masses $>1$
M$_\odot$, and another decreasing only for the last 20 My (solid
line), which gives $\Gamma=\Gamma_{infer}$ above 10 M$_\odot$ and
$\Gamma=-1.35$ for lower $M$. The fiducial slopes for inferred and
intrinsic IMFs are shown by the dashed lines.   If these examples
apply to the field regions of the LMC and SMC for masses above
$10$ M$_\odot$, for example, then the SFR had to drop there by
factors of 63, 9.2, and 1.9, respectively, during the last
$\sim20$ My. For $\Gamma=-1.8$ between 7 and 35 M$_\odot$ observed
by Parker et al., the required SFR drop is 1.6 in 47 My, the
lifetime of a 7 M$_\odot$ star.  Such drops would not be
unreasonable if the field regions were sites of dense clouds and
star formation $\sim50-20$ My ago, and these clouds have since
dispersed and moved elsewhere by the pressures of young stars.

Models of star formation in the LMC suggest the overall rate has
decreased steadily for the last $\sim$Gy, following a large burst
at about that time in the past (Bertelli et al. 1992; Westerlund,
Linde, \& Lyngå 1995; Vallenari et al. 1996; Gallagher et al.
1996). While the details of this decline are uncertain, this is
the type of SFH that could explain a moderately steep
($\Gamma\sim-2$) field IMF down to $\sim2$ M$_\odot$.  For the
three assumed values of $\Gamma_{infer}=-4,$ $-3$, and $-2$ in
Figure 6, the SFR drops over the last Gy by factors of 3780, 112,
and 5.0. Gallagher et al. allow for a factor of $\sim3$ decrease
in the SFR over the last Gy.  Thus a declining SFR in the LMC
could explain the moderately steep field IMF inferred by Parker et
al. (1998) and others, even down to $\sim2$ M$_\odot$ if needed.
That is, the factor of $\sim5$ drop in SFR over 1 Gy suggested for
our $\Gamma=-2$ case in Figure 6 is about the same as the factor
of $\sim3$ drop observed by Gallagher et al. for 1 Gy. More
precisely, for the $\Gamma_{infer}=-1.8$ obtained by Parker et
al., the SFR has to drop by a factor of 2.6 over 1 Gy; this is
almost the same as the drop observed by Gallagher et al..
Gallagher et al. also rule out a drop in the SFR by more than a
factor 10 in the last 0.1 Gy. This is consistent with the Massey
et al. result of $\Gamma_{infer}\sim-4$ for $M>40$ M$_\odot$
because that requires a drop of a factor of $\sim3.3$ over the
last 3.7 My, the lifetime of a 40 M$_\odot$ star according to
equation 2.

Other models for steep field $\Gamma$'s include a greater
dispersal from OB associations of low-mass stars compared to high
mass stars resulting from the longer lives of the low mass stars
(Elmegreen 1997, 1999; Hoopes, Walterbos \& Bothun 2001; Tremonti
et al. 2002), a dependence of maximum stellar mass on cluster mass
that is either the result of statistical sampling effects (Kroupa
\& Weidner 2003) or in addition to these effects (Elmegreen 1999),
the evaporation of cluster halos that have a steep IMF from birth
(e.g., de Grijs et al. 2002), and inadequate corrections for
low-mass background stars (Parker et al. 2001). Field
contamination by runaway stars that are the minor companions of
binaries can steepen the IMF too. All of these processes, in
addition to decelerating star formation, can steepen both the PDMF
and the inferred field IMF while allowing a somewhat shallow,
cluster-like, intrinsic IMF.

\section{Conclusions}
Variations in the star formation rate over times from
$\sim2\times10^6$ to $10^9$ yr, as expected from observations of
our own and other galaxies, can produce variations in the slope of
the present day mass function that will be misinterpreted as
variations in the slope of the IMF if the star formation history
is assumed to vary more smoothly or not at all.   The resulting
IMFs can have false flattenings or steepenings at masses whose
main sequence age is comparable to the time since the star
formation event.  IMF steepening can be large if star formation
turns off suddenly.

Applications of these results to the field regions of the LMC and
SMC appear promising. There the inferred IMF is relatively steep
for intermediate mass stars, $\Gamma\sim-1.8$ compared to a
Salpeter slope of $-1.35$, and even steeper for high mass stars,
$\Gamma\sim-4$ for $M>40$ M$_\odot$. We found that the
$\Gamma\sim-1.8$ observation by Parker et al. (1998) between 7 and
35 M$_\odot$ could result from a decline in the average star
formation rate by a factor of 1.6 over the last 47 My if the
intrinsic IMF is the Salpeter function. Similarly, the
$\Gamma\sim-4$ observation by Massey (2002) for $M>40$ M$_\odot$
could result from a decline by a factor of 3.3 over the last 3.7
My. If $\Gamma=-1.8$ all the way down to 2 M$_\odot$, then the
required decline would be a factor of 2.6 over the last 1 Gy.
These declines would have had to occur throughout the entire LMC
and SMC if the inferred steep IMFs are the same in all field
regions.  If this is the case, then the field stars could be from
dispersed clusters where the intrinsic IMF was relatively shallow
at birth.

Our results apply also to IMF determinations of young clusters
with a substantial pre-main sequence population if the IMF is
estimated from the luminosity function. That problem is more
complicated than the cases modeled here because there is no
one-to-one relation between mass and luminosity for pre-main
sequence stars.  In order to obtain a reliable IMF for pre-main
sequence stars, the SFH should be determined simultaneously with
the IMF using theoretical isochrones in the H-R diagram -- a
procedure that has been used by several groups but can still
introduce systematic errors if there are uncertainties in the
isochrones and effective temperature scale. Older clusters should
not have a problem with their inferred IMFs if the main sequence
lifetime of the most massive star is longer than the length of
time during which the cluster formed.

Applications of SFH variations to the field IMF are additionally
complicated by the fact that the present local field population
consists of stars with a wide range of ages that originated in
diverse parts of the Galaxy having different SFHs. The range of
birth sites increases with decreasing mass because main sequence
lifetime and mean age increase with decreasing mass. The range in
birthplace distance is the result of stellar orbit diffusion,
which is rapid in the azimuthal direction, increasing with the
square root of the stellar age (Wielen 1977; see Rocha-Pinto et
al. 2004 for a review of processes). Radial mixing by stellar
waves was recently studied by Sellwood \& Binney (2002). Because
there is a wide range of birth distances for stars spanning the
inferred field IMF, many different SFR histories are involved. We
neglected this effect here.  For stars that are too young for
significant orbit diffusion (ages less than a few $10^8$ yr), this
neglect is equivalent to assuming that all SFR variations are
coherent over spatial scales larger than the stellar age times the
average velocity dispersion. In principle, the age spread of solar
neighborhood stars with a given mass can significantly affect the
inferred IMF if each star comes from a region with a different
SFH. For this reason it will be difficult to infer any intrinsic
IMF from field stars without detailed models of their migration
into the local neighborhood and their spatial and temporal
distribution of formation rates.

Helpful comments by the referee are gratefully acknowledged.  JMS
acknowledges support through NASA ATP Grant NAG5-13280.

\begin{figure}\plotone{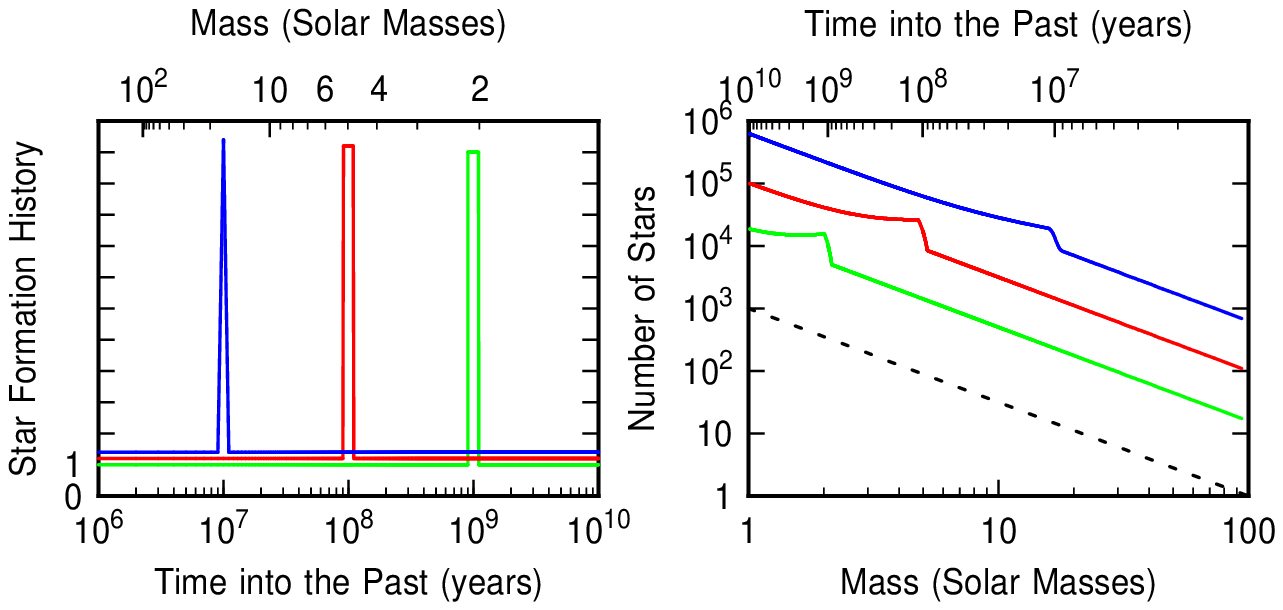}
\caption{(left) Three sample star formation histories consist of a
constant rate with a burst at various times. Present time is to
the left. Successive plots are shifted upward for clarity. The
masses whose main sequence lifetimes correspond to the ages on the
abscissa are indicated on the top axis. The corresponding present
day mass functions derived from these SFHs were divided by the
main-sequence age as a function of mass to give the inferred IMFs
on the right. These would be the IMFs inferred for the region if
the star formation rate were erroneously taken to be constant. The
inferred IMFs have features at masses where the main sequence ages
equal the burst times.  The main sequence lifetime for each mass
is indicated on the top axis. The dotted line has the intrinsic
IMF slope of $\Gamma=-1.5$.}\end{figure}

\begin{figure}\plotone{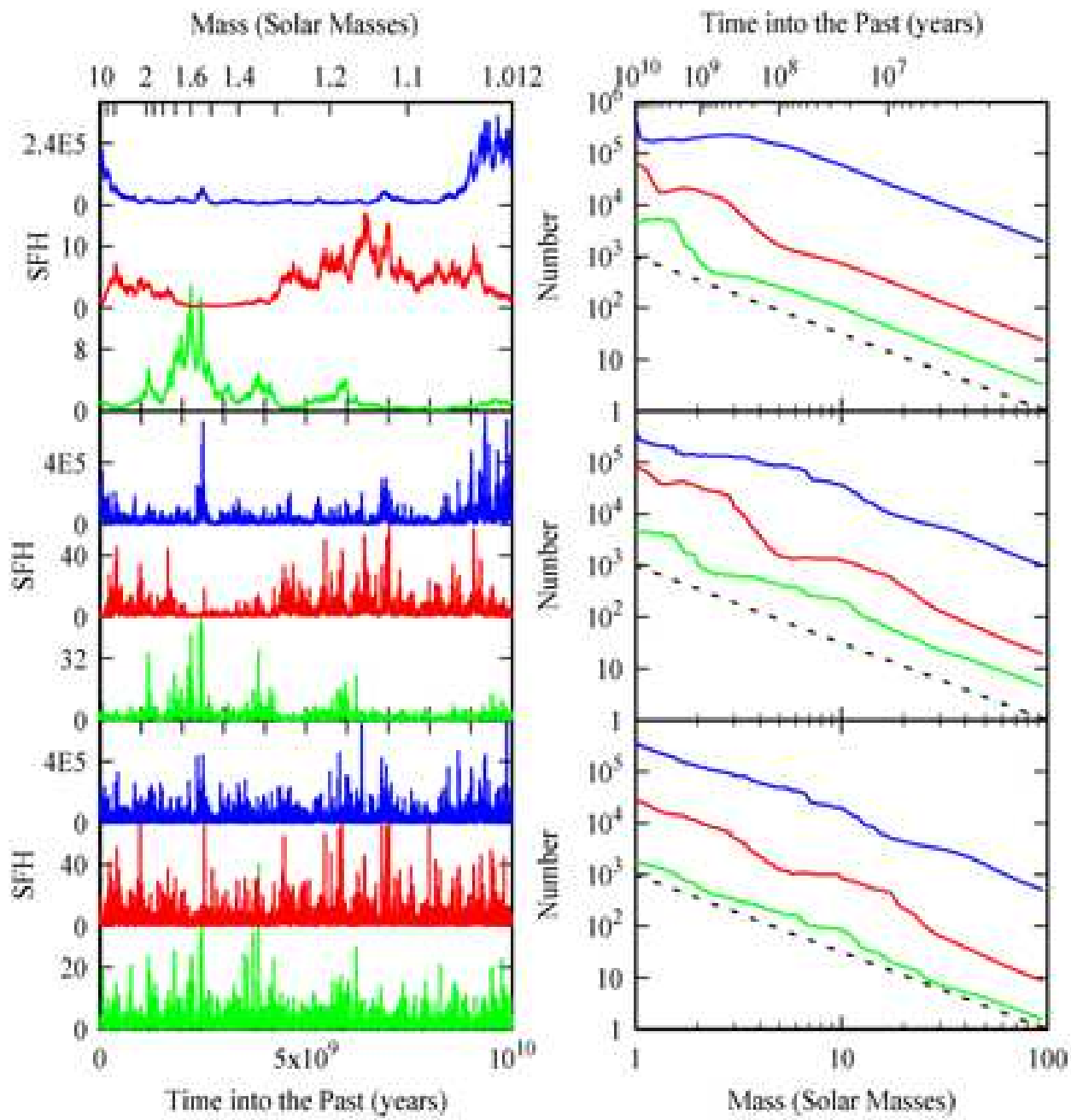}
\caption{Star formation histories (SFH) that have log-normal pdf's
and approximate power law power spectra (left) and their
associated inferred IMFs (right). Successive plots are shifted
upward for clarity. The dotted line on the right has the intrinsic
IMF slope of $\Gamma=-1.5$. The histories were found by generating
noise in Fourier space, multiplying the noise by $k^{-\beta}$,
taking the inverse Fourier transform, normalizing the resulting
Gaussian distribution function of values to an rms dispersion of 1
and an average of 0.1, 1, and 10 (for the lower to upper curves in
each panel, respectively), and then exponentiating the result to
get a log-normal distribution. Values of $\beta=0.25,$ 0.5, and 1
are shown at the bottom, middle and top, corresponding to power
spectra of with slopes of $-0.5$, $-1$, and $-2$.}\end{figure}

\begin{figure}\plotone{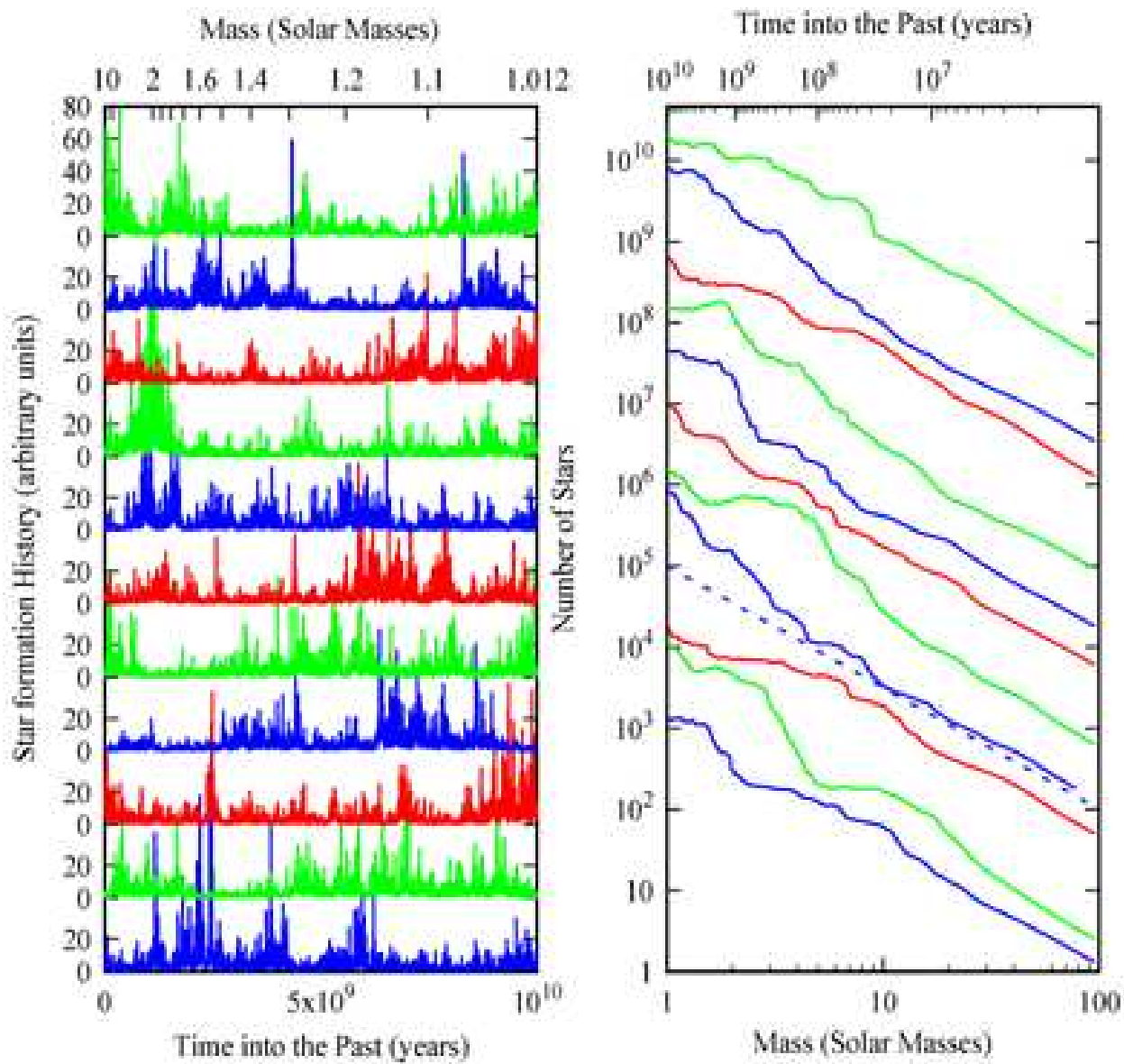}
\caption{Eleven random examples of stochastic star formation
histories (SFH) with log-normal pdf's and approximate power law
power spectra (left) and their associated inferred IMFs (right).
The SFH parameters are the same as in one of the cases in Fig. 2:
$\beta=0.5$ and an average value before exponentiation equal to 1.
Thus, the middle example in the middle panel of Fig. 2 is
reproduced here as the second curve up from the bottom in both
left and right-hand panels. Successive plots are shifted upward
for clarity. The dotted line on the right has the intrinsic IMF
slope of $\Gamma=-1.5$.}\end{figure}

\begin{figure}\plotone{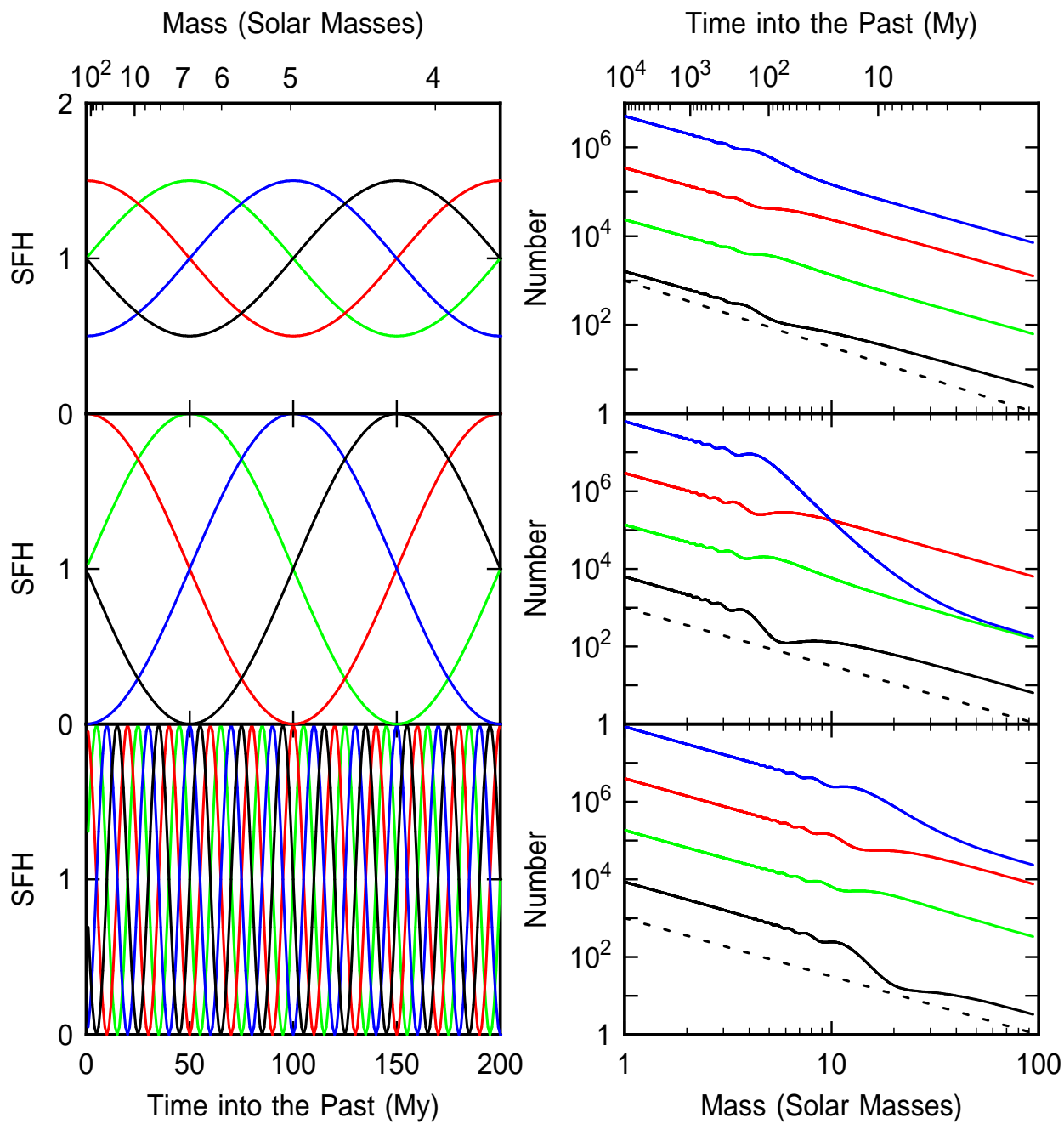}
\caption{Star formation histories that are sine functions with
periods of 20 My on the bottom and 200 My in the middle and top
panels, along with their inferred IMFs. The minimum values of the
SFH  are set equal to 0 in the two bottom panels and 0.5 in the
top panel. The amplitude of the inferred IMF feature increases
with increasing relative amplitude of the SFR variation. The local
slope of the IMF depends strongly on the phase of the SFH. The
dotted line on the right has the intrinsic IMF slope of
$\Gamma=-1.5$.}\end{figure}

\begin{figure}\plotone{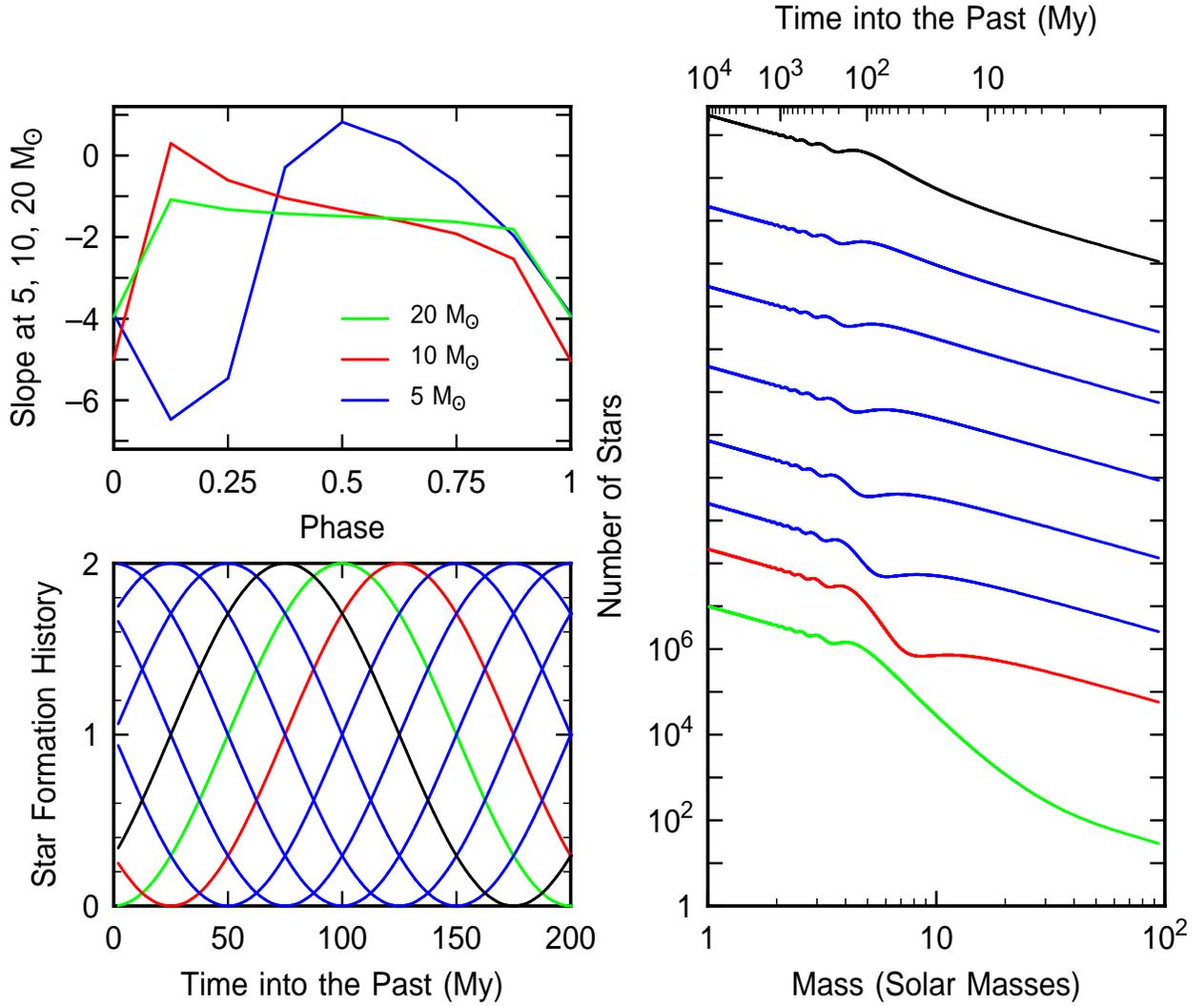} \caption{Star
formation histories are shown in the bottom left panel for eight
sine functions with varying phases. The corresponding inferred
IMFs are on the right in ascending order (i.e., the SFH that
starts at 0 for recent time is shown as the lower IMF). The top
left panel shows the inferred IMF slopes $\Gamma_{infer}$ at three
stellar masses, 5 M$_\odot$, 10 M$_\odot$, and 20 M$_\odot$. The
slope at each mass is steeper when the SFH has just decreased
prior to the time corresponding to the main sequence lifetime at
that mass.}\end{figure}

\begin{figure}\plotone{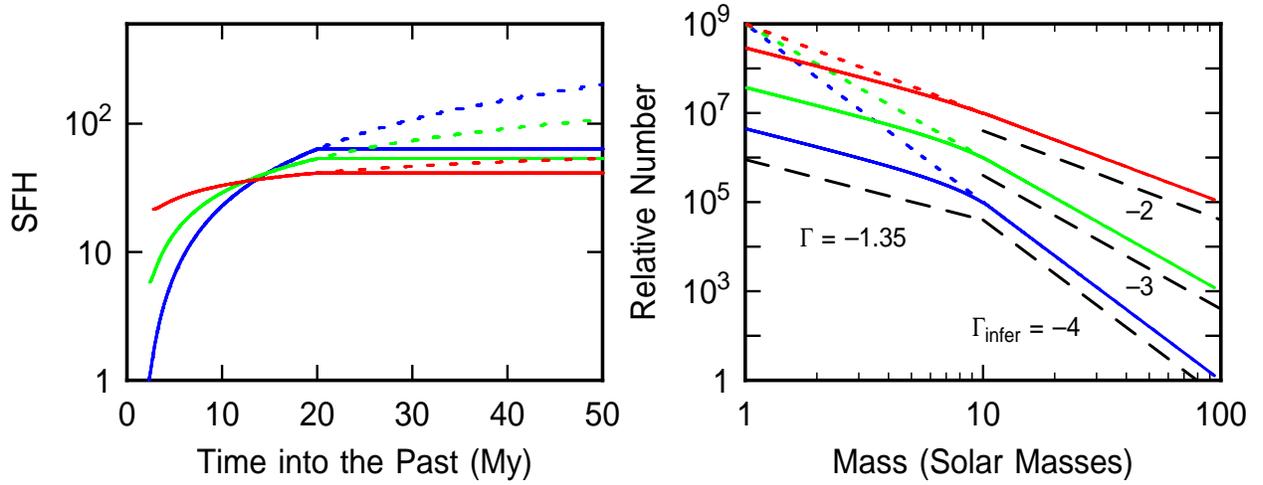}
\caption{Star formation histories (left) that give an inferred IMF
with a slope of $\Gamma_{infer}=-4$ (blue, with the biggest
decline), $-3$, and $-2$ (red, with the smallest decline), as
suggested for the field regions of the Large and Small Magellanic
clouds. The dashed lines give these slopes down to 1 M$_\odot$ and
the solid lines give them to 10 M$_\odot$.  The resulting IMFs are
shown on the right. The intrinsic IMF slope chosen to illustrate
these cases is $\Gamma = -1.35$.}\end{figure}
\end{document}